\documentclass[10pt,twocolumn]{IEEEtran}
\usepackage{multicol}
 
\usepackage{color}
\usepackage{xcolor}
\usepackage{soul}
\usepackage{amsmath}
\usepackage{graphics}
\usepackage{setspace}
\usepackage{cite}
\usepackage{latexsym}
\usepackage{float}
\usepackage{epsfig}
\usepackage{multirow}
\usepackage{cite,cases,url}
\usepackage{amssymb}
\usepackage{graphicx}
\usepackage{epstopdf}
\usepackage{fancyhdr}
\usepackage{afterpage}
\usepackage{balance}
\usepackage{subcaption}
\usepackage{enumerate}
\usepackage{array}
\usepackage{algorithmic}
\usepackage{algorithm}
\usepackage{enumitem}
\usepackage{dblfloatfix}
\usepackage[normalem]{ulem}
\useunder{\uline}{\ul}{}

\newcolumntype{L}{>{\centering\arraybackslash}m{5cm}}
\newcolumntype{K}{>{\centering\arraybackslash}m{6cm}}
\newcolumntype{P}{>{\centering\arraybackslash}m{2.3cm}}
\newcolumntype{M}{>{\raggedright\arraybackslash}m{2cm}}
\newcolumntype{N}{>{\raggedright\arraybackslash}m{2.5cm}}

\fancypagestyle{firstpage}
{
    \fancyhead[L]{\footnotesize This article has been accepted for publication in the IEEE Communications Standards Magazine. Please cite it as A. S. Abdalla and V. Marojevic, ''Communications Standards for Unmanned Aircraft Systems: The 3GPP Perspective and Research Drivers,'' IEEE Communications Standards Magazine, June 2021. } 
    \fancyhead[R]{\footnotesize \thepage}
    \cfoot{}
    
}

\begin{document}

\title{Communications  
Standards for Unmanned Aircraft Systems: The 3GPP Perspective and Research Drivers}

\author{\IEEEauthorblockN{Aly Sabri Abdalla and Vuk Marojevic} \\
\IEEEauthorblockA{Dept. Electrical and Computer Engineering, Mississippi State University,
Mississippi State, MS, USA\\
asa298@msstate.edu, vuk.marojevic@msstate.edu}

\vspace{-8mm}
}

\maketitle

\thispagestyle{firstpage}
\begin{abstract}
An unmanned aircraft system (UAS) consists of an unmanned aerial vehicle (UAV) and its controller which use radios to communicate. While the remote controller (RC) is traditionally operated by a person who is maintaining visual line of sight with the UAV it controls, the trend is moving towards long-range control and autonomous operation. To enable this, reliable and widely available wireless connectivity is needed because it is the only way to manually control a UAV or take control of an autonomous UAV flight. 
This article surveys the ongoing Third Generation Partnership Project (3GPP) standardization activities for enabling networked UASs. In particular, we present the requirements, envisaged architecture and services to be offered to/by UAVs and RCs, which will communicate with one another, with the UAS Traffic Management (UTM), and with other users through 
cellular networks. 
Critical research directions relate to security and spectrum coexistence, among others. We identify major R\&D platforms that will 
drive the standardization of cellular communications networks and applications. 
\end{abstract}

\IEEEpeerreviewmaketitle
\section{Introduction}
\label{sec:intro}

\begin{figure*}[t]
    \centering
    \includegraphics[width=1\textwidth]{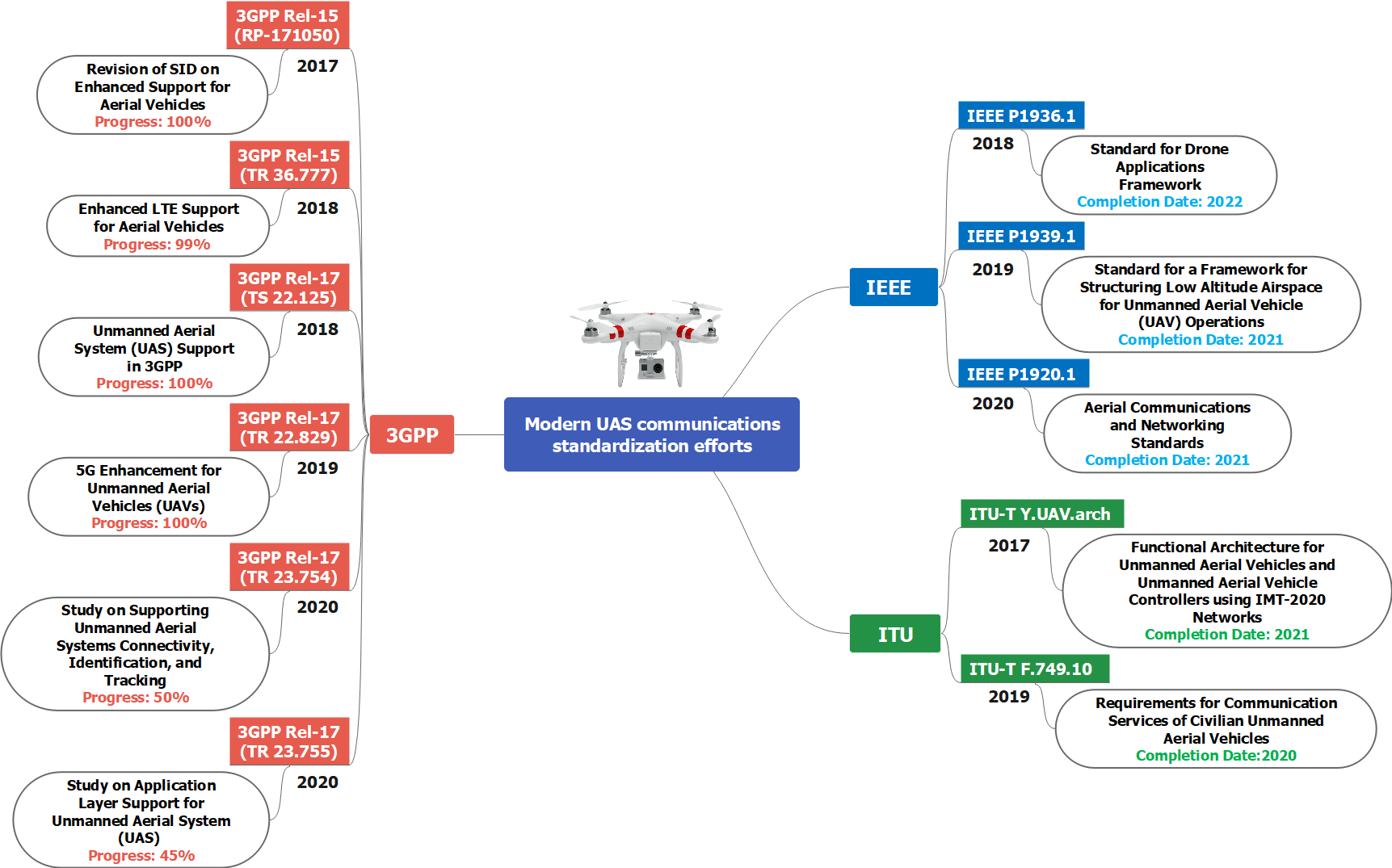}
    \caption{
    Recent standardization efforts for UAS communications with completion percentages as of November 2020.}
    \label{fig:Figure1}
    \vspace{-3mm}
\end{figure*}

Unmanned aerial vehicles (UAVs) have been successfully used for the delivery of goods in humanitarian, medical and commercial contexts. 
UAVs are also considered critical 
to support  
public safety missions in different domains, including search and rescue operations and disaster management.  
Another valuable use case for UAVs is precision agriculture where UAVs can be used to manage and monitor crops, detect weed, and collect ground sensors data (moisture, soil, etc.) in areas where there is no network or no business case for deploying one. 
Their low deployment cost, high maneuverability, and ability to operate in dangerous or hostile environments are some of the reasons for choosing UAVs over other technologies. 

An unmanned aircraft system, or UAS, is commonly considered as a UAV and its controller. The UAV controller can directly control the UAV from a nearby operator using a remote controller (RC), or indirectly through a terrestrial or satellite communications network. 
The growing market for commercial UAVs has led to global research and development (R\&D) on communications and networking technology, protocols and applications supporting UAV operation. 
Industry 
has started  
the integration process of UAVs into 4G and 5G cellular networks.

The Third Generation Partnership Project (3GPP), an industry-driven consortium that is standardizing cellular networks, has been active in identifying the requirements, technologies and protocols for aerial communications.  
{An early  
study to use enhanced long-term evolution (LTE) support for aerial vehicles  
successfully concluded 
at the 3GPP RAN\#76 meeting in 2017 with the Report RP-171050  
that led to the 3GPP Release 15 (Rel-15) Technical Report TR~36.777. 
This report 
identifies the necessary enhancements to LTE 
to optimize the network performance when serving UAVs. 
The first item that is discussed 
is that of developing a robust communications framework for UAS services that sustain high levels of safety and authorized use, and that embody the requirements and dynamics of the airspace regulation. 
As a result, the 3GPP 
Technical Specification TS~22.125 
establishes the requirements for providing  
UAS services through 3GPP networks. 

The 3GPP TR~22.829, published in 2019, 
identifies several UAV-enabled applications and use cases that shall be supported by 5G networks and 
points out the necessary communications and networking performance improvements. 
During 2020, the work items related to UAS communications 
in Rel-17 focus on two major aspects: the network infrastructure and procedures to support the connectivity, identification, and tracking of UAVs (TR~23.754) and 
the application architecture 
to support efficient UAS operations (TR~23.755). 

}

Other standardization bodies, including the IEEE and the International Telecommunication Union (ITU), are active in this area. 
Different IEEE standards committees address air-to-air communications standards for self-organized ad hoc aerial networks, low altitude UAV traffic management and UAV applications \cite{IEEE1920}. 
The ITU aligns with the 3GPP vision and, among others, establishes a functional architecture for UAVs and RCs, acting as user terminals that access IMT-2020 networks \cite{R3}. These 
activities are captured in Fig. 1.

The objective of this article is to 
present the recent 3GPP 
standardization efforts for the UAS integration 
into emerging cellular networks in such a way to explain the motivation behind the communications requirements and proposed solutions and to identifying the remaining research challenges and opportunities.
Prior work has illustrated the connectivity opportunities, gaps and interference problems~\cite{R1, R4}. 
Reference~\cite{R2} introduces the 3GPP TR~36.777, which discusses air-to-ground interference with UAVs connected to terrestrial LTE networks. 
Zeng et al.~\cite{R3} and Yang et al.~\cite{R5} share  their vision on 5G networks supporting a  diverse set of  drone-enabled applications and discuss some of the early ITU and 3GPP standardization efforts, respectively. 
This paper provides more insights into the ongoing 3GPP standardization efforts for integrating UASs into 
cellular networks with respect to UAV identification, command and control (C2), broadband applications and services, and coexistence with ground users. We 
dissect the 3GPP standardization efforts from the perspectives of the communications requirements (Section II) and the corresponding  
solutions (Section III).
 We  
identify 
important challenges for future research, development in Section IV.  
Section V introduces three major  
experimental R\&D 
projects. 
Section VI derives the concluding remarks with a brief technology outlook.

\section{UAS Identification, Control and Wireless Service 
Requirements}
\label{sec:requirements}

The 3GPP has established certain requirements for effective communications with UAVs. First, the UAS nodes 
need to register and authenticate with the network and provide regular status notifications. This is done by remote identification.
Then 
C2 and other wireless services can be facilitated through the 
cellular network. 
We describe the requirements here and the proposed technological solutions 
in Section III. 

\subsection{Remote Identification of UAS Nodes}

For the awareness, safety and efficient management of airspace operations, an aircraft needs to provide identification and regular presence updates to the control towers and airspace management system. For UASs this means that both the UAV and its controller need to register and regularly confirm their presence to the supporting network, which facilitates this information to the UAS Traffic Management (UTM). The UTM is a 
centralized flight management system developed and maintained by government 
authorities 
to provide aviation support and control for 
UAV operations at low altitude.  
It  
supports
strategic aerial deconfliction between coexisting operations, airspace authorization, in-flight reroute, remote identification, and flight intent sharing with other authorizations~\cite{UTMML}. 
The main component of the UTM is the UAS service supplier  
that works as a bridge between UAS operators and air navigation service providers.  
Supplemental data service providers further support UAS operations and an authorization entity ensures authorized access to these services. 

The 
identification requirements and related procedures are summarized as follows: 
\begin{enumerate}
    \item The UAS nodes  
    register to the network and provide information that enables 
    associating the UAV and its controller as UAS node pairs to the UTM.
    
    \item The transport network 
    forwards the UAV originating data---the UAV identity, its capability as a user equipment (UE),  
    make and model, serial number, take-off weight, position, owner information, 
    take-off location, mission type, flight data, and operation status---to the UTM. 
    
    \item The transport network 
    forwards the UAV controller data---the controller identity, its capability as a UE, controller position, owner, operator and pilot information, 
    and flight plan---to the UTM. The UAV controller is typically a ground control station or an RC. 
    
    \item The UAS nodes  
    use the network to transfer any additional 
    data that may be relevant  
    to the UTM after  
    successful 
    authentication and authorization.  
    
    \item The network 
    forwards live position information updates of the UAV and its controller to the UTM. Both UAS nodes therefore 
    transmit beacons at regular time intervals. 
    
    \item The network  
    gathers the UAS identification and subscription information for discriminating between a UAS capable UE and non-UAS capable, or regular, UEs. 
    A UAS capable UE---UAV-UE or controller-UE---is a radio transceiver that 
    communicates with the UTM as well as with other network 
    entities and application servers.   
\end{enumerate}

\begin{table*}[ht]
\centering
\caption{KPIs for the four C2 modes according to 3GPP TS 22.125, V17.1.0.}

{\begin{tabular}{|p{3.75cm}|p{1.2cm}|p{1.48cm}|p{1.75cm}|p{1.05cm}|p{1.3cm}|}
\hline
\centering
\textbf{Control Mode} & \textbf{Message\newline Interval} &  \textbf{Max UAV\newline Speed} &\textbf{Message Size} & \textbf{Latency} & \textbf{Positive ACK} 
\\ \hline
\vspace{-0.05 in}
{Steer to waypoints\newline (UAV terminated) } &
\vspace{-0.05 in}
$<$ \text{1 s}
&
\vspace{-0.05 in}
300 km/h
&
\vspace{-0.05 in}
100 byte
&
\vspace{-0.05 in}
1 s
&
\vspace{-0.05 in}
Required
\\ \hline
\vspace{-0.05 in}
{Steer to waypoints\newline (UAV originated) } &
\vspace{-0.05 in}
 \text{1 s}
&
\vspace{-0.05 in}
300 km/h
&
\vspace{-0.05 in}
84-140 byte
&
\vspace{-0.05 in}
1 s
&
\vspace{-0.05 in}
Not required
\\ \hline
\vspace{-0.05 in}
{Direct stick steering \newline (UAV terminated) } &
\vspace{-0.05 in}
 \text{40 ms}
&
\vspace{-0.05 in}
60 km/h
&
\vspace{-0.05 in}
24 byte
&
\vspace{-0.05 in}
40 ms
&
\vspace{-0.05 in}
Required
\\ \hline
\vspace{-0.05 in}
{Direct stick steering \newline (UAV originated) } &
\vspace{-0.05 in}
 \text{40 ms}
&
\vspace{-0.05 in}
60 km/h
&
\vspace{-0.05 in}
84-140 byte
&
\vspace{-0.05 in}
40 ms
&
\vspace{-0.05 in}
Not required
\\ \hline
\vspace{-0.05 in}
{Automatic flight on UTM \newline (UAV terminated) } &
\vspace{-0.05 in}
 \text{1 s}
&
\vspace{-0.05 in}
300 km/h
&
\vspace{-0.05 in}
$<$ \text{10K byte}
&
\vspace{-0.05 in}
5 s
&
\vspace{-0.05 in}
Required
\\ \hline
\vspace{-0.05 in}
{Automatic flight on UTM \newline (UAV originated) } &
\vspace{-0.05 in}
 \text{1 s}
&
\vspace{-0.05 in}
300 km/h
&
\vspace{-0.05 in}
1500 byte
&
\vspace{-0.05 in}
5 s
&
\vspace{-0.05 in}
Required
\\ \hline
\vspace{-0.05 in}
{Approaching Autonomous Navigation Infrastructure \newline (UAV terminated) } &
\vspace{-0.05 in}
 \text{500 ms}
&
\vspace{-0.05 in}
50 km/h
&
\vspace{-0.05 in}
4k byte
&
\vspace{-0.05 in}
10 ms
&
\vspace{-0.05 in}
Required
\\ \hline
\vspace{-0.05 in}
{Approaching Autonomous Navigation Infrastructure \newline (UAV originated) } &
\vspace{-0.05 in}
 \text{500 ms}
&
\vspace{-0.05 in}
\centering
-
&
\vspace{-0.05 in}
4k byte
&
\vspace{-0.05 in}
140 ms
&
\vspace{-0.05 in}
Required
\\ \hline
\end{tabular}%
}
\label{tab:KPI2}
\end{table*}

\subsection{C2 Communications} 

There are four C2 modes to control UAV flight operations. 

\begin{itemize}
    \item \textbf{Steer to waypoints:} 
    Pre-scheduled waypoints or other flight declarations are communicated from the UAV controller or UTM to the UAV.
    
    \item \textbf{Direct stick steering:} 
    The UAV controller directly communicates with the UAV in real time to provide flight directions through waypoints.

    \item \textbf{Automatic flight by UTM:} The UTM 
    provides this autonomous flight option 
    through an array of predefined 4D polygons. 
    The UAV feeds back periodic position reports for 
    flight tracking purposes. 
    
    \item \textbf{Approaching autonomous navigation infrastructure:} 
    The C2 infrastructure supports autonomous UAV flights and may provide updated flight instructions, such as the next waypoint, altitude and speed. It can also help with autonomous departing and landing operations. 
    \end{itemize}

Each one of these modes has specific requirements in terms of 
packet interval, message size, and end-to-end latency, among others.  
Table~\ref{tab:KPI2} captures 
the key
performance indicators (KPI) for these modes.  
The control links 
are bi-directional where messages  
are exchanged between the controller, the UTM, or both,
and the UAV. 
Correct packet reception acknowledgment (ACK) 
is required for all UAV terminated  
transmissions  
because they may contain critical UAV flight control instructions. 
The switch to another C2 mode during operation is possible if the requirements of the new mode can be met. 
The 3GPP recommends using video  
feedback for supporting direct stick steering with the following video parameters. 
\begin{itemize}
    \item  \textbf{Visual line of sight (VLOS) operation}: {
    2 Mbps data rate for a 480p video with 30 frames per second (fps) and 1 s latency.}
    \item \textbf{Beyond VLOS (BVLOS) operation}: {
    4 Mbps data rate for a 720p video with 30 fps and 140 ms latency.}
\end{itemize}

The BVLOS operation has more stringent video feedback requirements because the video is the only way for the pilot to follow the UAV flight.

\subsection{ 
Other Wireless 
Applications and Services}
For the full integration of the UAS into  
cellular networks, the 3GPP standardization  
is expected to support various UAV-assisted applications and services. 
Most of these are defined around video and photo delivery from the UAV for 
situational awareness, surveillance or entertainment purposes, among others.
Table~\ref{tab:KPI1} captures the fundamental 
services that UAVs are envisaged to provide and the corresponding KPIs. 
The currently specified services and KPIs will be leveraged and extended to support future engineering, scientific, emergency, and other commercial and non-commercial applications and use cases. 
\subsection{Coexistence with Terrestrial Users}
\begin{figure}[t]
    \centering
    \includegraphics[width=0.49\textwidth]{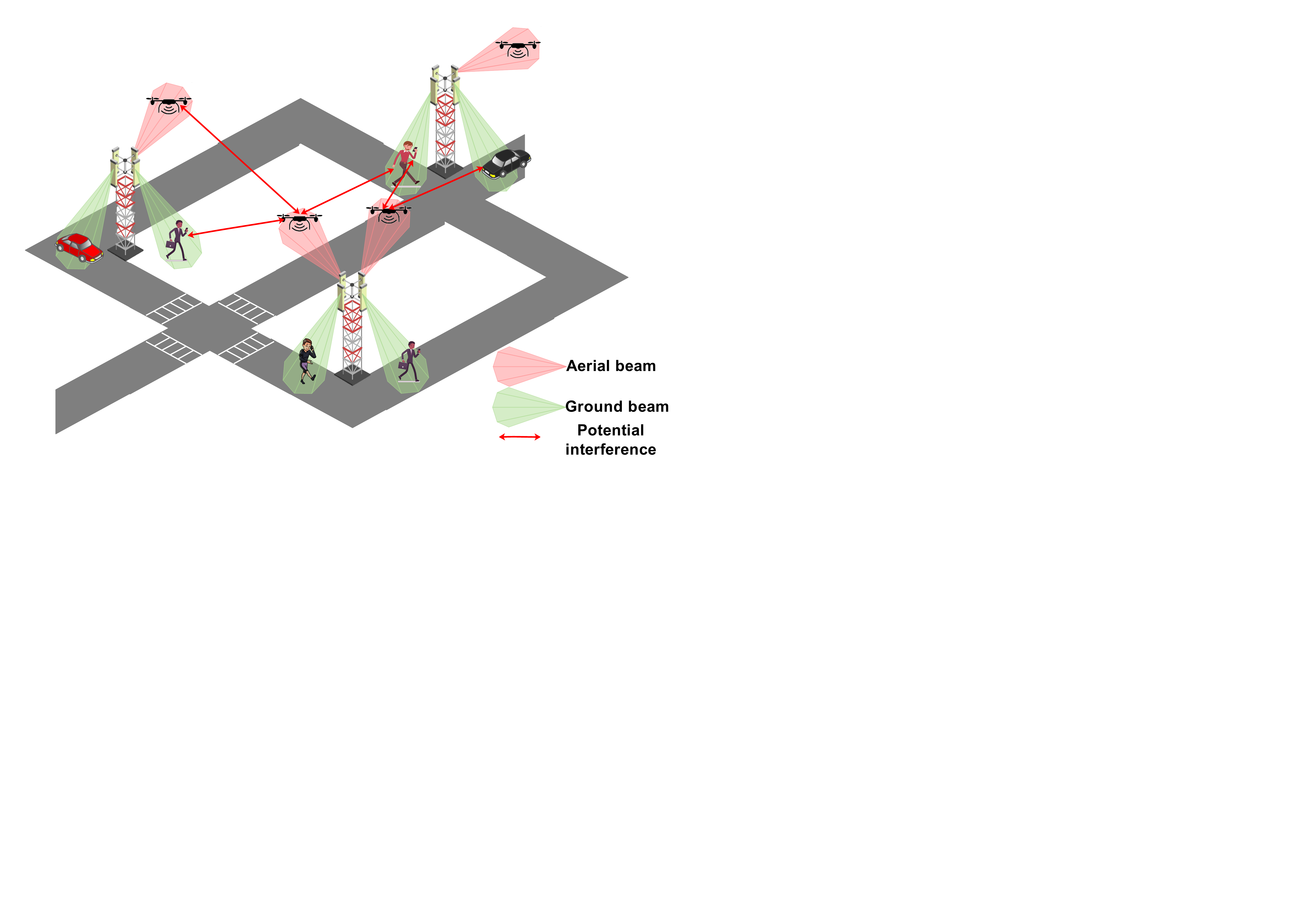}
    \caption{
    Cellular communications infrastructure with beam-based downlink transmission supporting both ground and UAV users.
    }
    \label{fig:FigureX}
    \vspace{-5mm}
\end{figure}

The coexistence of aerial and ground UEs and the resultant interference that can occur on both the uplink and the downlink have been studied 
by the 3GPP   
for Release-15. 
This study has found that the aerial UEs will see more cells and potentially cause interference to 
a 
number of neighboring cells, especially at higher altitudes. 
Legacy cellular networks take advantage of the natural attenuation on the ground  
and have developed a hexagonal cellular grid model for initial cell tower planning and user association. Intercell interference typically involves only two or three cells and mainly affects users at cell edges if not properly handled. Aerial users 
will have dominant line of sight (LoS) links to multiple base stations and ground UEs in different terrestrial cells. 
As a result, the ground-UE uplink capacity in multiple cells may be impacted by a single UAV-UE uplink transmission. Similarly, the UAV-UE may receive multiple base station transmissions that interfere with one another, unless they are coordinated. 
In general, the network performance may suffer with increasing resource utilization per ground or aerial user. 
Therefore, a combination of coordinated resource scheduling, coordinated transmission, beamforming, or advanced interference mitigation solutions are needed to ensure coexistence between aerial and ground UEs while avoiding a fragmentation of spectrum.
Fig.~\ref{fig:FigureX}. illustrates the envisaged beam-based downlink transmissions and the potential radio frequency (RF) interference. 

\begin{figure}[t]
    \centering
    \includegraphics[width=0.45\textwidth]{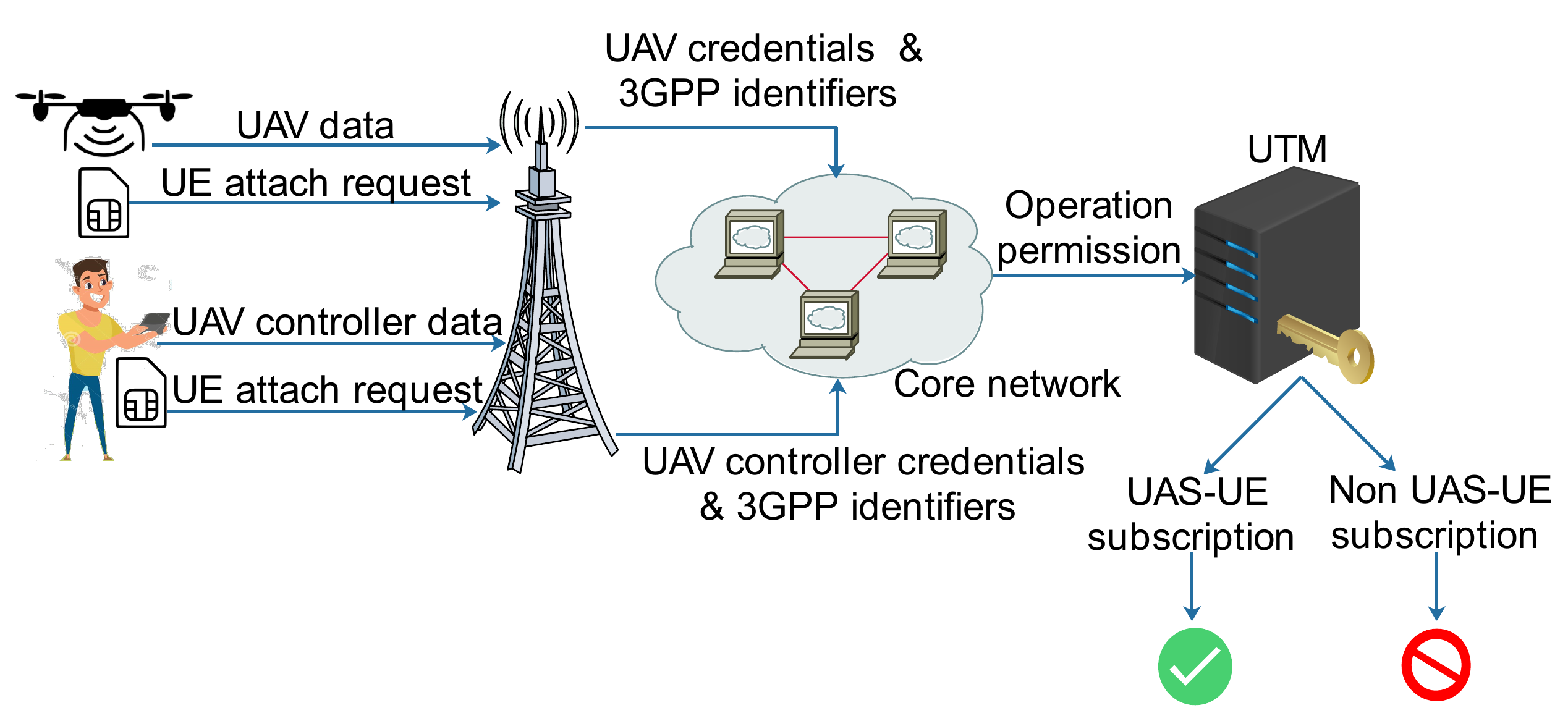}
    \caption{ The UAS authorization procedure.}
     \label{fig:Figure5}
    \vspace{-3mm}
\end{figure}


\begin{table*}[ht]
\centering
\caption{KPIs for UAV-assisted wireless services according to 3GPP TS 22.125, V17.1.0.}

{\begin{tabular}{|p{2.7cm}|p{2.38cm}|p{1.25cm}|p{1.5cm}|p{1.4cm}|p{1.3cm}|p{1.8cm}|p{2.4cm}|}
\hline
\centering
\textbf{Service} & \textbf{Uplink \newline Data rate} & \textbf{Uplink Latency} &\textbf{Downlink Data Rate} & \textbf{Downlink Latency} & \textbf{Height} &  \textbf{Service Area} &  \textbf{
Use Case}
\\ \hline
\vspace{-0.05 in}
8K video live broadcast &
\vspace{-0.05 in}
100 Mbps
&
\vspace{-0.05 in}
200 ms
&
\vspace{-0.05 in}
600 Kbps
&
\vspace{-0.05 in}
20 ms
&
\vspace{-0.05 in}
$<$ \text{100 m}
&
\vspace{-0.05 in}
Urban, scenic area
&
\vspace{-0.05 in}
Healthcare, virtual reality
\\ \hline
\vspace{-0.05 in}
{4*4K AI surveillance} &
\vspace{-0.05 in}
120 Mbps
&
\vspace{-0.05 in}
20 ms
&
\vspace{-0.05 in}
50 Mbps
&
\vspace{-0.05 in}
20 ms
&
\vspace{-0.05 in}
$<$ \text{200 m}
&
\vspace{-0.05 in}
Urban, rural area
&
\vspace{-0.05 in}

Face and object recognition, environment awareness, autonomous driving
\\ \hline
\vspace{-0.05 in}
{Remote UAV controller through HD video} &
\vspace{-0.05 in}
$\geq$ \text{25 Mbps}
&
\vspace{-0.05 in}
100 ms
&
\vspace{-0.05 in}
300 Kbps
&
\vspace{-0.05 in}
20 ms
&
\vspace{-0.05 in}
$<$ \text{300 m}
&
\vspace{-0.05 in}
Urban, rural area
&
\vspace{-0.05 in}
Non-LoS bidirectional UAV control
\\ \hline
\vspace{-0.05 in}
{Real-Time Video} &
\vspace{-0.05 in}
0.06 Mbps 
&
\vspace{-0.05 in}
100 ms
&
\vspace{-0.05 in}
\centering
-
&
\vspace{-0.05 in}
\centering
-
&
\vspace{-0.05 in}
\centering
-
&
\vspace{-0.05 in}
Urban, rural, countryside
&
\vspace{-0.05 in}
Entertainment 
with multiple  \newline camera views
\\ \hline
\vspace{-0.05 in}
{Video streaming} &
\vspace{-0.05 in}
4 Mbps$\to$  \text{720p} \newline
9 Mbps$\to$ \text{1080p}
&
\vspace{-0.05 in}
100 ms
&
\vspace{-0.05 in}
\centering
-
&
\vspace{-0.05 in}
\centering
-
&
\vspace{-0.05 in}
\centering
-
&
\vspace{-0.05 in}
Urban, rural, countryside
&
\vspace{-0.05 in}
Public safety, emergency surveillance,  law  enforcement
\\ \hline
\vspace{-0.05 in}
{Periodic still photos} &
\vspace{-0.05 in}
1 Mbps
&
\vspace{-0.05 in}
1 s
&
\vspace{-0.05 in}
\centering
-
&
\vspace{-0.05 in}
\centering
-
&
\vspace{-0.05 in}
$<$ \text{120 m}
&
\vspace{-0.05 in}
Urban, rural, countryside
&
\vspace{-0.05 in}
Exploration, inspections, search \& rescue
\\ \hline
\end{tabular}
}
\label{tab:KPI1}
\end{table*}
\vspace{5mm}
\section{3GPP Communications Solutions}
\label{sec:contribution}

The communications network that serves UASs needs to facilitate robust and scalable communications and networking services.  
The cellular network that can offer such services will be able to  
support safe and efficient UAS operation, as well as added functionalities 
enabled by the broad coverage that cellular networks provide. 

\subsection{Remote Identification of UAS Nodes}
The 3GPP 
relies on the UTM for managing the UAS node authentication and credential information.  
That is, all required authorization and credential checks for the UAV 
and its controller to gain 
access to the network and authorized services will be performed by the UTM. From the point of view of the 3GPP,  
the aerial communications subscription information and the radio capability are the main parameters.

Fig.~\ref{fig:Figure5} shows the authorization procedure for a UAS 
to be registered and authorized by the 3GPP network. Once the UAS nodes are authenticated and authorized to use the network for UAV control and other applications and to connect to the UTM, additional authentication mechanisms at the application layer must be triggered to activate the UTM services, such as UAV collision avoidance  
and live position tracking. 
If the 
UAS attach request does not match the aerial communications subscription at the UTM, the request will be declined. 
An operation request of a UAV may 
be denied  
even if the identities or credentials are successfully verified. 
The remote identification of UAS nodes 
will provide information about the deployed 
UAVs, active operators, the specific geographical area, and so forth, which can aid in detecting suspicious behavior or prevent nuisance complaints.    
\subsection{C2 Communications}
A C2 link
carries the UAV flight control commands from the ground control station or RC to the UAV.  
It also carries telemetry data from the UAV to the 
controller to facilitate efficient operation.  
The UAV has a number of distinct flight functions that can be managed by a ground operator through the UAV controller. 
In addition,  
C2 connectivity can be established between the UAV and the UTM for other air traffic management 
operations. 
\begin{figure}[t]
    \centering
    \includegraphics[width=0.49\textwidth]{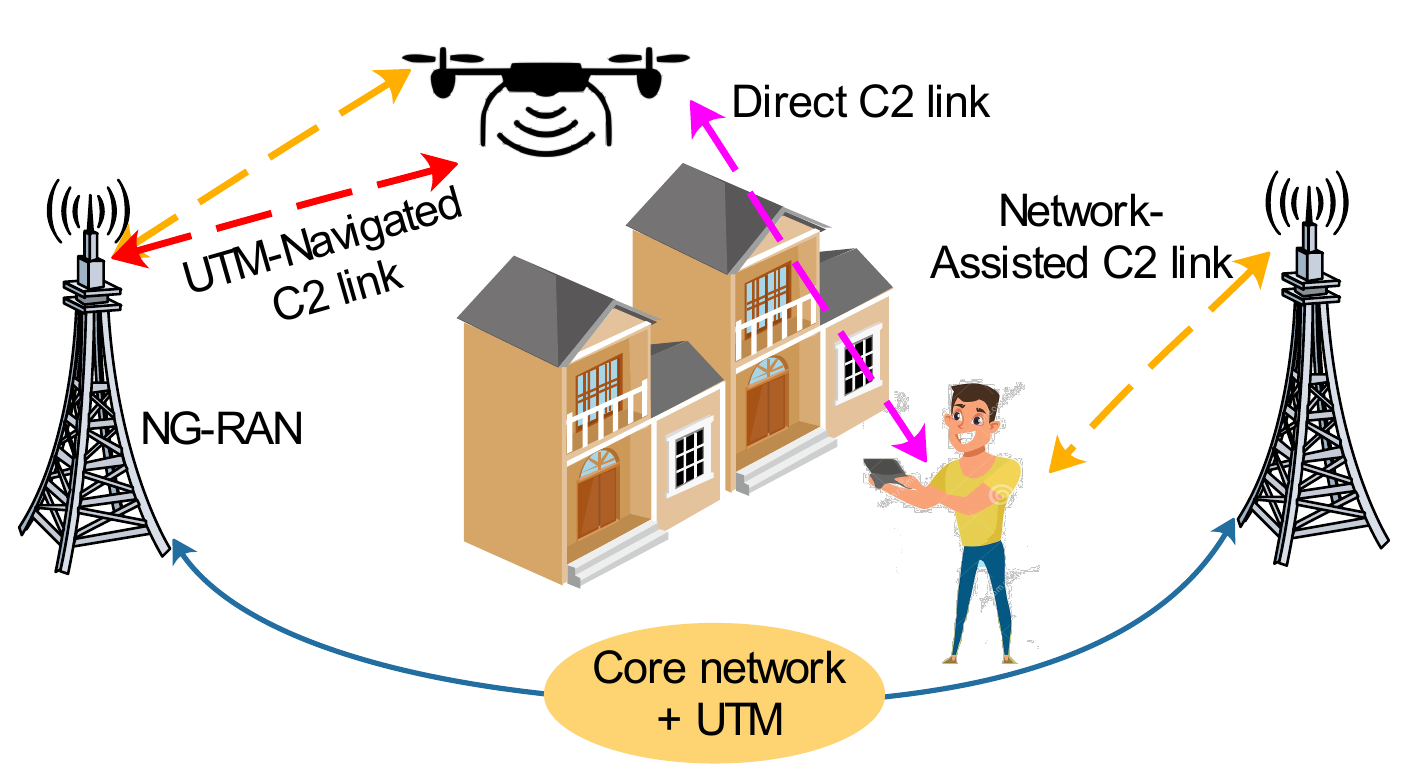}
    \caption{The three C2 communications paths. }
    \label{fig:Figure2}
    \vspace{-5mm}
\end{figure}
{The UTM plays a vital role for air traffic management  
and its deployment  
can be 
centralized or decentralized.  
It should be able to transmit the path information and flight clearance to the UAV during the entire flight. 
The communications system should therefore maintain a latency below 500 ms, needed to deliver any route modification notification 
to the UAV. 
Short-range broadcasts from UAVs for avoiding  
aerial vehicle collisions also need to be supported by the network.
} 

The cellular  
network can be used to assist different UAS communications services as a transport network. 
For the sake of reliability  
and service quality, more than one 
connection can be established adding redundant C2 
links. 
Fig.~\ref{fig:Figure2} illustrates the envisaged C2 and UTM support through the cellular network. 
It shows three control paths:

\begin{itemize}
    \item \textbf{Direct C2 communications:} 
    This direct connection between the UAV and its controller will be activated only if both nodes 
    are authenticated and registered to the 
    network. 
    The control commands  
    will be carried over the resource that may be configured and scheduled by the 
    network, taking into 
    consideration the C2 communications requirements. 
    \item \textbf{Network-assisted C2 communications:} Unicast C2  
    links can be established between the cellular 
    network and the UAV as well as between the 
    network and the UAV controller. 
    The UAV and its controller 
    do not need to be close to one other and may register to the 
    network through  
    different radio access networks (RANs). 
    \item \textbf{UTM-navigated C2 communications:}  
    This type of 
    connectivity is used by the UTM among others to monitor the UAV flight status, 
    to deliver flight 
    path updates, to keeping track of the UAV navigation, and to provide flight navigation orders whenever needed~\cite{UTMML}.
    Typically, most of the UTM-navigated C2 links will be established indirectly over a 3GPP  
    network server. 
\end{itemize}

Is should be noted 
that the steer to waypoints and the direct stick steering packets can be delivered using the direct or network-assisted C2 connectivity, whereas the other two fundamental C2 modes described in Section II.B 
are to be provided by the UTM-navigated C2 network.

\subsection{RAN Support} 
\begin{figure}[t]
    \centering
    \includegraphics[width=0.48\textwidth]{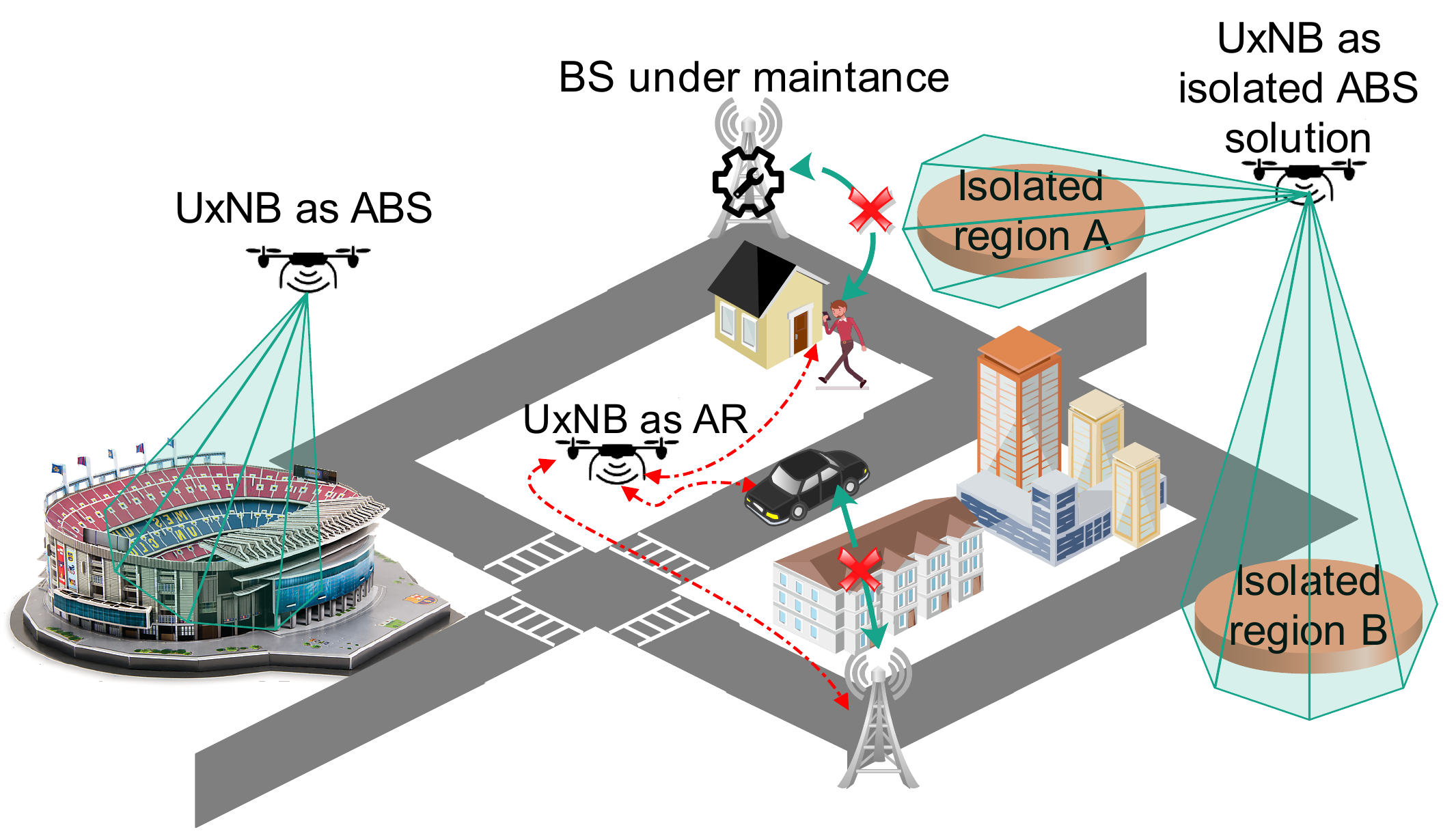}
    \caption{UxNB deployment scenarios.}
    \label{fig:Figure3}
    \vspace{-3mm}
\end{figure}

The deployment of modern radios as UAV payloads
allows to establish broadband data links for numerous applications, including aerial support nodes for terrestrial RANs. 
The motivation for this is the 
ability of UAVs to navigate in the three-dimensional space and hover in place with minimal 
restrictions and at low deployment and maintenance costs. Such use of UAVs can support the different generations of cellular networks. 
The 3GPP calls such an aerial radio node the on-board radio access node (UxNB). 
The UxNB can be used to extend the coverage or increase the capacity of the cellular network. It allows rapid deployment in disaster and emergency use cases, e.g. for supporting evacuation where communications infrastructure may be lost or for providing capacity on demand at crowded events. 
The UxNB node can implement an aerial base station (ABS), an aerial relay (AR), or an isolated 
ABS solution, as illustrated in Fig.~\ref{fig:Figure3}. 
The UxNB flies to the designated area where wireless services are needed and then hovers there while providing wireless connectivity to ground users.

Before the UxNB starts its mission, it needs to be authorized by the network management system 
and configured 
as a function of the specific operation, objectives and internal and external parameters. The UxNB can then serve a specific 
geographical area for a given time, using certain spectral resources. 

The flight time of the UxNB is effected by various factors that can shorten its operation. 
Those factors include the UxNB size, on-board battery, cargo, aerodynamics, and authorized airspace regulations. Therefore, it has been recommended that the 3GPP systems assisting the aerial missions of the UxNB nodes 
support the close monitoring and reporting of the UxNB status. This includes 
the UAV's power consumption, position, trajectory, mission, and the environmental conditions. This continuous monitoring and reporting will facilitate achieving the required QoS for the given mission for an extended period, which can be accomplished by  
managing dynamic UxNB replacements.

Another emerging application of the UxNB is 
to provide local routing and proximity-based services (ProSe), as introduced in the 3GPP~TS 24.334. 
UxNB can, for example, establish ProSe group communications among ground users located in remote areas without network coverage.  
For such a scenario, the deployed node will act as an ABS and provide a RAN with or without a backhaul connection. 
In the latter case, the UxNB will allow communications 
among the  
UEs in the area served by the UxNB  
by establishing 
an internal IP network. 
On the other hand, the isolated ABS 
with a backhaul link will support 
routing of selected IP traffic 
to external IP networks, such as the Internet. 

The huge potentials and benefits of deploying UxNB nodes in the sky motivated researchers and industry to define and standardize 
a set of services. 
Those services 
will provide the means to minimize the power consumption of the UxNB by optimizing the operational parameters, UxNB path planning, and service delivery. In addition, it is important to consider and monitor the radio frequency interference  
among the deployed UxNB nodes and terrestrial cellular users for effective network operation. This is important because of the generally strong LoS conditions and the fact that the deployment of a UAV at a specific location may not have extensive field data collected beforehand.  

As has been mentioned before, the UxNB's greatest weakness is its short operational time and, therefore, it is important  
to optimize the management of UxNB node replacement for uninterrupted service provisioning. 

\vspace{-2mm}
\subsection{Coexistence with Terrestrial Users}
The 3GPP Technical Report 
TR 36.777 identifies various 
solutions to the RF interference problem when integrating aerial cellular users. For example, the base stations may employ multiple antennas and support full-dimension multiple-input multiple-output 
communications for transmit/receive beamforming 
to minimize 
inter-cell interference 
and aerial-ground interference. Null-steering can further enhance performance, especially for ground users~\cite{FDMIMO1}. 
In addition, the use of directional antennas  for beamforming at the aerial UEs will lower the RF footprint of UAV originating  
transmissions to better isolate ground and aerial signaling.
Uplink interference can be further controlled using open loop and closed loop power control for both aerial and ground UEs, as well as new scheduling, admission, and congestion control mechanisms that take UAV and UE locations into account.

Aerial-ground UE coexistence mechanisms will need to be devised as function of the UE and base station capabilities, as well as the density of cellular-connected UAVs. The wireless industry that is developing and operating networks and UEs and driving 3GPP standards is expected to define coexistence requirement and candidate technologies with support for backward compatibility.

\section{Research Challenges and Opportunities}
\label{sec:Challenges}

Fundamental challenges are still to be tackled for the full integration of UAVs into emerging cellular networks. 
In what follows, we identify what we believe are among the most compelling research directions.

\begin{itemize}
    \item \textbf{
    Location-Based Services:} The integration of UAVs into cellular networks requires defining a new category of devices, 
    subscriptions 
    and differentiated services.  
    The 3GPP introduces the UE height parameter to differentiate between a UAV and a legacy UE. 
   There are several instances where more accurate and GPS independent localization is needed. There are three main approaches: 
   visual, inertial navigation system (INS), and wireless signaling. Visual techniques use on-board cameras.  
    INS-based solutions monitor the orientation, location and velocity of the UAV. 
    However, the performance of the inertial measurement unit has been shown to degrade with flight time~\cite{INS}. 
    The third group relies on RF signaling and channel characterization. 
    Research that combines on-board sensor data with the help of wireless protocol and infrastructure support to provide robust localization services will support many aspects of UAV communications and networking, including those discussed below.
    
    \vspace{+1 mm}
    \item \textbf{Multi-UAV  
    Management:} The short battery life and limited payload capacity of small UAVs are 
    their major weakness. 
    Moreover, UAVs will be flying at different speeds and altitudes, or be hovering, which will lead to frequent changes of the aerial network topology. 
    As a result, most of the current or proposed deployments of UAVs are anchoring on the use of multiple vehicles or 
    swarms
    that can replace one another  
    without service disruption.  
    Therefore, it is essential to have a low-latency, proactive, and scalable management system to control the handover of missions and transition of roles. 
    5G 
    technologies, such as network slicing and  softwarization offered by software-defined networking and network function virtualization, can be leveraged for managing 
    UAV networks~\cite{survey_Soft}. While these technologies allow flexible network use and custom network services, how to ensure isolation, scalability, and QoS guarantees for UAV control and data communications will be an important R\&D theme in the coming years.
     This includes 
     analyzing the UAV originating RF transmission footprint and evaluating centralized versus distributed scheduling, congestion control, and routing mechanisms~\cite{FANET_Routing}, which must be adaptable, scalable, and of low latency. 
     There is research interest in designing and prototyping self-optimizing UAV networks \cite{Selfoptimiz_Drones}. 
     Dynamic decision making solutions that may trigger changes to the network configuration, UAV positions, or trajectories, among others, 
     need to be further explored  
     to enable efficient aerial networking.
     
     \vspace{+1 mm}
    \item \textbf{Security:} As UAVs show huge potential for supporting wireless services,  
    wireless security becomes of critical concern. 
    The aerial nodes are vulnerable to 
    attacks, such as unauthorized access and control, eavesdropping of 
    data transferred between UAVs and ground control stations, jamming of GPS signals or UAV communications links, and location and identity spoofing attacks. 
    Therefore, providing secure and reliable wireless links and different levels of integrity and privacy protection 
    mechanisms must be sustained by the standards as mandatory features and enforced in practice. The research community has investigated various options to provide a secure, resilient and self-configurable framework for UAS communications. These options include blockchain solutions as a defense system against UAV network softwarization attacks. Cross-layer authentication and mutual authentication mechanisms have been proposed for improving the confidentiality and integrity of UAV communications, specifically at the application layer. In addition, physical layer security techniques, such as artificial noise transmission and relaying, can be leveraged with UAVs to assist terrestrial networks~\cite{UAV_WirelessMag}. 
    A forth UxNB mode may become that of an aerial reconfigurable intelligent surface (RIS), which allows to change the propagation environment with passive RF elements to, for instance, avoid jamming or eavesdropping~\cite{RIS}.          
    \vspace{+1 mm}
     \item \textbf{Spectrum Coexistence:}
     The RF interference emerging from cellular-connected UAVs can be mitigated by leveraging the LoS-dominant air-to-ground channels for RF sensing~\cite{SpectSharing}. It is then possible to coordinate the 
transmissions between the UAV and the terrestrial users as a cognitive radio based solution. 
Moreover, RISs deployed on building walls or even carried by UAVs can help reflect beams intelligently to allow lower power transmissions or create interference-free regions~\cite{RIS}. 
Different waveform configurations, dynamic channel access and adaptive scheduling can be driven by machine learning algorithms, facilitated by data collected during testing and predeployment. While certain types of data, such as C2 signals, are time critical, other data may be scheduled for transmission as a function of UAV position, heading, RF and network congestion, among others. Massive deployments of next generation base stations, remote radio heads, relays or RISs on buildings in coordination with the use of UxNBs will help reducing the transmission powers to effectively increase capacity for both the ground and aerial users coexisting with one another. These techniques require experimental research to collect data and devise the most effective coexistence mechanisms, not only between aerial and terrestrial cellular network users, but also with other active and passive radio services in unlicensed and shared spectrum where next generation wireless networks will increasingly operate.
\end{itemize}

\section{
R\&D Platforms}
\label{sec:projects}

While UAS communications research and standardization are still in their early stages, important 
R\&D 
projects and testbeds are being established. These  
are necessary  
to study the performance requirements of UAVs and evaluate the technology and protocol solutions to support the various use cases in real-life scenarios. Repeatable experimental results obtained in controlled, yet production-like environments, will in return accelerate advancing communications 
standards. 
\begin{itemize}

\item \textbf{AERPAW [https://aerpaw.org]}: The Aerial Experimentation and Research Platform for Advanced Wireless (AERPAW) is being built in the US since 2019 under the Platforms for Advanced Wireless Research (PAWR) program. 
AERPAW is a unique large-scale testbed enabling experimentation with advanced wireless technology and systems for UAVs. 
The goal of AERPAW is to support global 5G and Beyond 5G wireless research on connected 3D mobility, spectrum agility and security, and 3D network topology, among others. 
AERPAW will therefore offer access to commercial-grade 5G technology and networks as well as to software radios that can be programmed to implement many different waveforms and protocols. These radios will be deployed on several fixed nodes and be available as payloads for UAVs. 
An emulator and sandbox will enable development and pre-testbed deployment of new radios, networks and experiments.
\vspace{+1 mm}
\item \textbf{5G!Drones [https://5gdrones.eu]:} In June 2019, a three-year European project named 5G!Drones
has kicked offed. Academia has joined forces with industry and includes network operators and research centers for testing UAV use cases over 5G networks. The use cases that are considered in this project includes: UAV traffic management, public safety, and situational awareness. The trails aim to validate the 
ability to support aerial services and provide feedback that can be used to improve the performance of 5G systems for the selected use cases. 5G!Drones is part of Phase 3 of the 5G 
Public Private Partnership (5GPPP) projects funded by the European Commission.
\vspace{+1 mm}
\item \textbf{5G-DIVE [https://5g-dive.eu]:} In October 2019, a collaboration  
of 
vendors, service providers, network
operators, small or medium-sized enterprises (SMEs), academic and research centers from 
the EU and Taiwan established the 5G-DIVE project.  
5G-DIVE plans to perform field and real-life tests of different 5G technologies to ensure technical merits and business value proposition are fulfilled, before advancing those technologies to higher levels. The 
trails mainly focus on 
Industry 4.0 use cases, such as digital twin app and real time video analysis for zero defect manufacturing. The second trails focus on 
autonomous drone scouting 
involving drone fleet
navigation and intelligent processing of images captured by the
drones. 
5G-DIVE is funded by the European Union through the H2020 Program. 
\end{itemize}
\section{Conclusions}
\label{sec:conclusions}

{UAS technology 
is providing innovative solutions to support various applications in the public, private, and military sectors. 
The benefits come not only from their mission-oriented use, but also as a general platform for a number of different purposes, including network support. 
This has been realized by the 3GPP that is refining the 
terrestrial cellular network for supporting 3D mobile users and networks. 
This article 
surveys the  
3GPP 
standardization efforts for networking UAS nodes and integrating them into cellular networks as end users or RAN support nodes. 
We introduce the 
initial 3GPP work items meant for 4G LTE and cover the recent standardization activities for 5G networks until mid 2020. 
We discuss the UAS node registration, C2, and wireless services requirements for UAVs. 
The article moreover discusses the emerging and future use cases of 
UASs, research challenges and opportunities, 
and 
major experimentation platforms 
for closing the gap between fundamental research, standardization, development, and deployment. 

Because cellular networks were designed and optimized for terrestrial users and specific communications services, the cellular network architecture will change to better serve 3D users, aerial support nodes, and new use cases. 
Application oriented KPIs, new terrestrial and ad hoc networking infrastructure proposals, functional splits, operational modes, and alternative signaling and protocol solutions for serving UAVs in their different roles will be the topic of research 
for years to come.}
This research, when demonstrated on an experimental platform in a production-like environment, will enable future standardization studies 
and drive new use cases for advanced wireless technology to spur innovation at the intersection of communications, networking, computing, control, and applications. 

\section*{Acknowledgement}
A.~S.~Abdalla and 
V.~Marojevic are supported in part by the National Science Foundation under grant numbers CNS-1939334 and ECCS-2030291.

\balance

\bibliographystyle{IEEEtran}
\bibliography{Ref}

\section*{Biographies}
\small
\noindent
\textbf{Aly Sabri Abdalla} (asa298@msstate.edu)
is a PhD candidate in the Department of Electrical and Computer Engineering at Mississippi State University, Starkville, MS, USA. His research interests are on scheduling, congestion control and wireless security for vehicular ad-hoc and UAV networks.

\vspace{0.2cm}
\noindent
\textbf{Vuk Marojevic} (vuk.marojevic@msstate.edu) is an associate professor in electrical and computer engineering at Mississippi State University, Starkville, MS, USA. His research interests include resource management, vehicle-to-everything communications and wireless security with application to cellular communications, mission-critical networks, and unmanned aircraft systems.

\end{document}